\documentstyle[preprint,aps,epsfig]{revtex}
\tightenlines
\begin{document}
\title{Direct capture astrophysical $S$ factors at low energy}
\author{B.~K.~Jennings, S. Karataglidis, and T. D. Shoppa}
\address{TRIUMF, 4004 Wesbrook Mall, Vancouver, B.C., Canada, V6T 2A3}
\date{\today}
\maketitle
\begin{abstract}
We investigate the energy dependence of the astrophysical $S$ factors
for the reactions $^7$Be$(p,\gamma)^8$B, the primary source of
high-energy solar neutrinos in the solar $pp$ chain, and
$^{16}$O($p,\gamma$)$^{17}$F, an important reaction in the CNO cycle.
Both of these reactions have predicted $S$ factors which rise at low
energies; we find the source of this behavior to be a pole in the $S$
factor at a center-of-mass energy $E = -E_{\rm B}$, the point where
the energy of the emitted photon vanishes.  The pole arises from a
divergence of the radial integrals.
\end{abstract}

The $^7$Be($p,\gamma$)$^8$B reaction, at center-of-mass energies 
$E$ near 20~keV, plays an important role in the 
production of solar neutrinos
\cite{bahcall}. The neutrinos from the subsequent decay of $^8$B
provide the high energy neutrinos to which many solar neutrino
detectors are sensitive. The cross section for this reaction is
conventionally expressed in terms of the $S_{17}$ factor, where
the $S$ factor is defined in terms of the cross section $\sigma$
by
\begin{equation}
S(E) = {\sigma(E) E} 
\exp (2\pi\eta(E))\;,
\end{equation}
where $\eta(E) = Z_1 Z_2 \alpha \sqrt{\mu c^2 / 2 E}$ is the
Sommerfeld parameter for nuclei of charges $Z_1, Z_2$ and reduced mass
$\mu$.  The exponential factor in the definition of $S$ removes the
rapid energy dependence of the cross section due to Coulomb repulsion
between the two nuclei.  In the stellar core the probability of
capture of protons by $^7$Be, obtained by folding the thermal
distribution of nuclei with the cross section, peaks at $\sim 20$~keV.
Because the cross section diminishes exponentially at low energies,
the only method of obtaining information about $S_{17}$ at those
energies is to extrapolate data taken at experimentally accessible
energies ($E > 100$~keV).

The $^{16}$O($p,\gamma$)$^{17}$F$^*$ ($\frac{1}{2}^+;\frac{1}{2}$,
0.498~MeV \cite{Ti93}) reaction, which occurs in the CNO
cycle, is of little importance for energy production in the sun but of
greater importance for hotter stars. As is the case for
$^7$Be($p,\gamma$)$^8$B, extrapolation of data taken at high energies
is necessary to obtain the $S$ factor at energies applicable in the
stellar core, $E \sim 25$~keV.

Direct capture calculations\cite{Christy} of these two reactions
\cite{Rolfs,williams,ROW} predict an
upturn in the $S$ factor at threshold. As the
capture in both reactions is primarily external, the $S$ factors at
astrophysical energies are determined by the product of the
spectroscopic factor, the asymptotic normalization of the final
(bound) state wave functions, and a purely Coulombic term. As the
spectroscopic factor is independent of energy, the energy dependence
of the $S$ factor, away from resonances, may be studied without
detailed knowledge of the nuclear structure.

In each reaction the weakly bound final state ($E_{\rm B}=137.5$~keV
\cite{Aj88} for $^8$B and $E_{\rm B}=105.2$~keV for the first excited
state of $^{17}$F \cite{Ti93}) causes the $S$ factor to rise as the
center-of-mass energy approaches 0.  In the case of $^{16}$O($p,
\gamma$)$^{17}$F$^{*}$ both the data and direct-capture calculations
\cite{Rolfs,Mo97} show clear evidence of this low-energy rise. The
$^{16}$O($p, \gamma$)$^{17}$F capture to the ground state shows no
such rise because the final state is more deeply bound and has higher
angular momentum.  For $^7$Be($p,\gamma$)$^8$B the upturn occurs below
the lowest experimental point so it is only observed in the
calculations.

Williams and Koonin \cite{williams} do an explicit expansion about zero
energy for $S_{17}$ and give the first two coefficients in a Taylor series for
the logarithmic derivative of $S_{17}$
as $-2.350$~MeV$^{-1}$ and $28.3$~MeV$^{-2}$.
Using these in a (1,1) Pad\'e approximant gives:  
\begin{equation}
S_{17}=\frac{(1+4.85 E)}{(1+7.20 E)}
\end{equation}
where $E$ is in MeV.  This Pad\'e approximant has a pole at $-139$~keV
which is very close to their bound state energy of 136~keV. Thus we
see that in the region of the threshold the bound state is important
and induces a pole. In fact, a Taylor series expansion would converge
only with a radius of the binding energy --- barely to the region that
is experimentally accessible. Hence any functional form for the
extrapolation to zero energy should contain the contribution from the
pole.

Following \cite{williams}, we write the  astrophysical $S$ factor 
as
\begin{equation}
S = C (I_0^2 + 2 I_2^2) E_\gamma^3 \left( J_{11} \beta_{11}^2 + 
  J_{12} \beta_{12}^2 \right) \frac{1}{1-e^{-2 \pi \eta}}\;,
\label{one}
\end{equation}
where
\begin{eqnarray}
I_L & = & \int^\infty_0 dr \; r^2 \; \psi_{iL}(r) \psi_f(r)/k \\
C   & = & \frac{5\pi}{9}  \frac{1}{(\hbar c)^3} (2 \pi \eta k) e^2 \mu^2
 \left( \frac{Z_1}{M_1} - \frac{Z_2}{M_2} \right)^2\;.
\end{eqnarray}
In Eq.~(\ref{one}), $J_{LS}$ is the spectroscopic factor for a given
angular momentum, $L$, and channel spin, $S$, $\beta_{LS}$ is the
asymptotic normalization of the bound state wave function, $E_\gamma$
is the photon energy, and $k$ is the momentum of the incident proton.
The final bound state wave function $\psi_{f}(r)$ is normalized
asymptotically to $\psi_{f}(r) = W_{\alpha,l}(\kappa r)/r$ while the
initial wave function reduces to the regular Coulomb wave function
divided by $\sqrt{2 \pi \eta}/(e^{2 \pi \eta}-1)$. The unusual choice
of normalizations is just to eliminate uninteresting factors from
quantities of interest. Most of those factors have been collected in
the coefficient $C$.
 
To investigate the behavior of the integrals in Eq.~(\ref{one}), we
first consider $\psi_{f}(r)=W_{\alpha,1}(\kappa r)/r$ for all radii
and take $\psi_{i0}(r)=F_0(k r)/(\sqrt{2 \pi \eta}/(e^{2 \pi
\eta}-1))$. The integral for the $s$-wave then becomes
\begin{equation}
I_0= \int^\infty_0 dr\;r W_{\alpha,1}(\kappa r)F_0(kr)/(k \sqrt{2 \pi \eta}/(e^{2 \pi \eta}-1)).
\end{equation}
At threshold, the integrands are peaked at large $r$: 40~fm for
$^7$Be($p,\gamma$)$^8$B, and 65~fm for
$^{16}$O($p,\gamma$)$^{17}$F$^*$. The tails of both integrands extend
well beyond 100 fm, and are, in each case, indicative of halo
states. The integral is smooth as $k$ passes through zero and diverges
as $k \rightarrow i\kappa$ ($E \rightarrow -E_{\rm B}$). The nature of
the divergence is determined by the asymptotic forms of the Coulomb
wave function and Whittaker function for large $r$. For large $r$ the
Whittaker function is proportional to $r^{- |\eta k|/\kappa} e^{-\kappa
r}$\cite{Christy} ($\eta k$ is independent of $k$). While above
threshold the Coulomb wave function oscillates at large radii, below
threshold it is exponentially growing and is proportional to
$r^{|\eta|} e^{|k| r}$. Thus the integrand approaches
\begin{equation}
r^{1-|\eta k|(1/\kappa-1/|k|)} \exp[-(\kappa-|k|)r] 
\end{equation}
for large $r$ and the integral diverges as 
\begin{equation}
I_0\propto1/(\kappa-|k|)^2 \propto 1/(E_{\rm B}+E)^2 = 1/E_\gamma^2\;.
\end{equation}
Since the
integrand diverges as $1/E_\gamma^2$ the leading term and first
correction term are both determined purely by the asymptotic behavior
of the wave functions. The first correction term is not simply
$1/E_\gamma$ but also involves logarithmic terms coming from the
$r^{-|\eta k|(1/\kappa-1/|k|)}$ factor.  The second correction term,
of order $E_\gamma^0$, is not determined purely by the asymptotic
value of wave function alone but also depends on the wave function at
finite $r$.

 From Eq.~(\ref{one}), we see that the quadratic divergence of $I_0$
gives rise to a simple pole in $S$ at $E_\gamma=0$. This suggests
writing the $S$ factor as a Laurent series: 
\begin{equation}
S = d_{-1} E_\gamma^{-1} + d_0 + d_1 E_\gamma +\ldots\label{laurent}
\end{equation} 
As before the coefficients of the first two terms, $d_{-1}$ and $d_0$,
are determined purely by the asymptotic forms of the wave functions
while the third coefficient, $d_1$, is also dependent on the short
range properties of the wave functions.

\begin{figure}
\epsfig{file=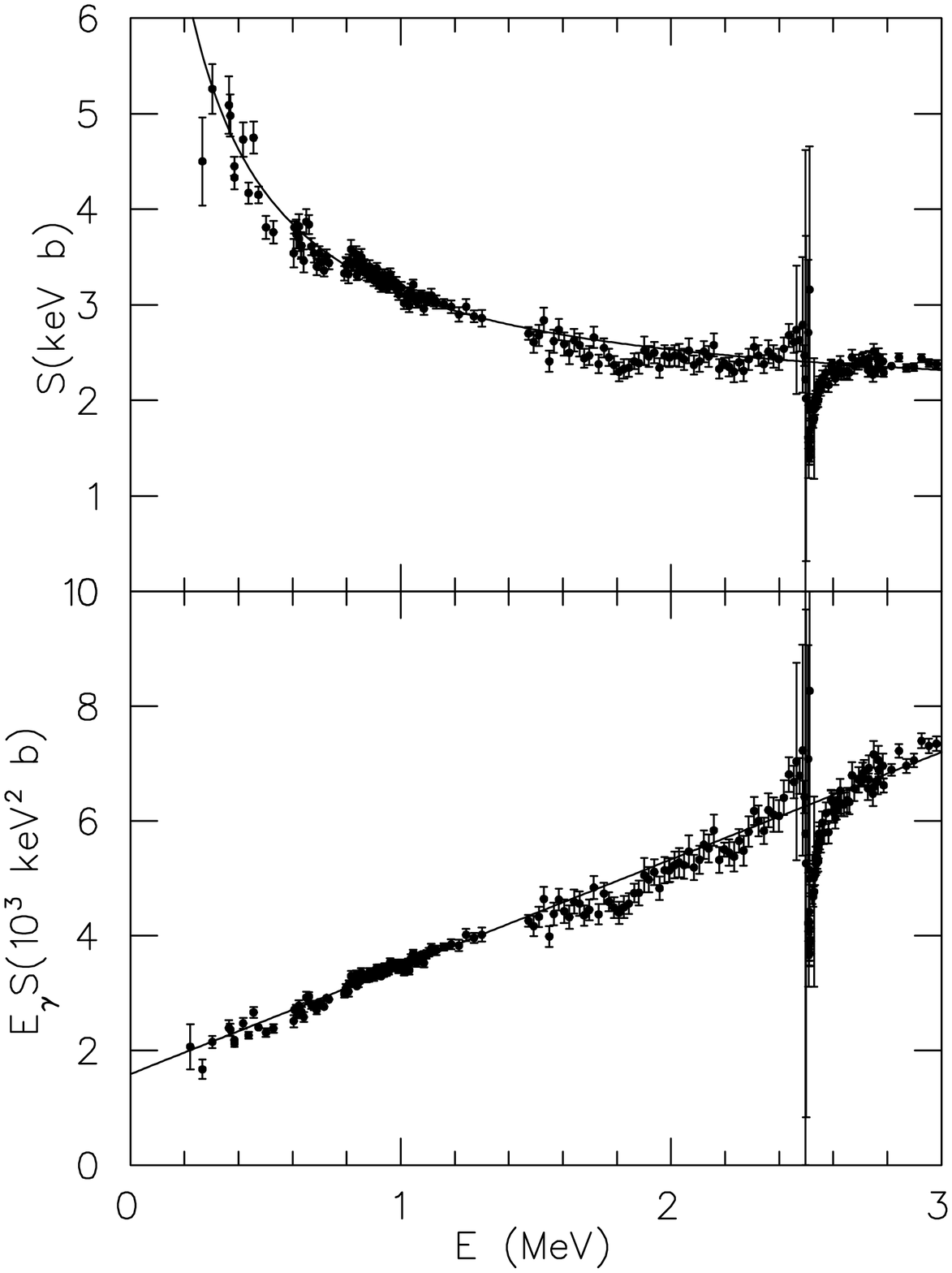,width=15cm}
\caption[]{Astrophysical $S$ factor (top) and $E_\gamma S$ (bottom) for
$^{16}$O($p,\gamma$)$^{17}$F$^{*}$. The data of Morlock {\em et al.}
\cite{Mo97} are compared to the fit as described in the text (solid
line).}
\label{s116}
\end{figure}

In Fig.~\ref{s116}, we present the data of Morlock {\em et al.}
\cite{Mo97} for the $^{16}$O($p,\gamma$)$^{17}$F$^{\ast}$ $S$ factor
(top) and for the product $E_\gamma S$ (bottom).  In the top panel,
the energy dependence of the $S$ factor is well approximated by the
form:
\begin{equation}
S = n \frac{1+c_1 E}{E_\gamma}
  = n \frac{1+c_1 E}{E+E_{\rm B}}
\label{form}
\end{equation}
where the constants $c_1$ and $n$ are determined by the straight line
fit to $E_\gamma S$ shown in the bottom panel.  The numerical values
are given in Table~\ref{param}.  There is remarkable agreement with
the data except near the resonance at 2.504~MeV.  Eq.~(\ref{form}) is
a convenient form for fitting experimental data and is motivated by
both the Pad\'e approximant and Eq.~(\ref{laurent}).

\begin{figure}
\epsfig{file=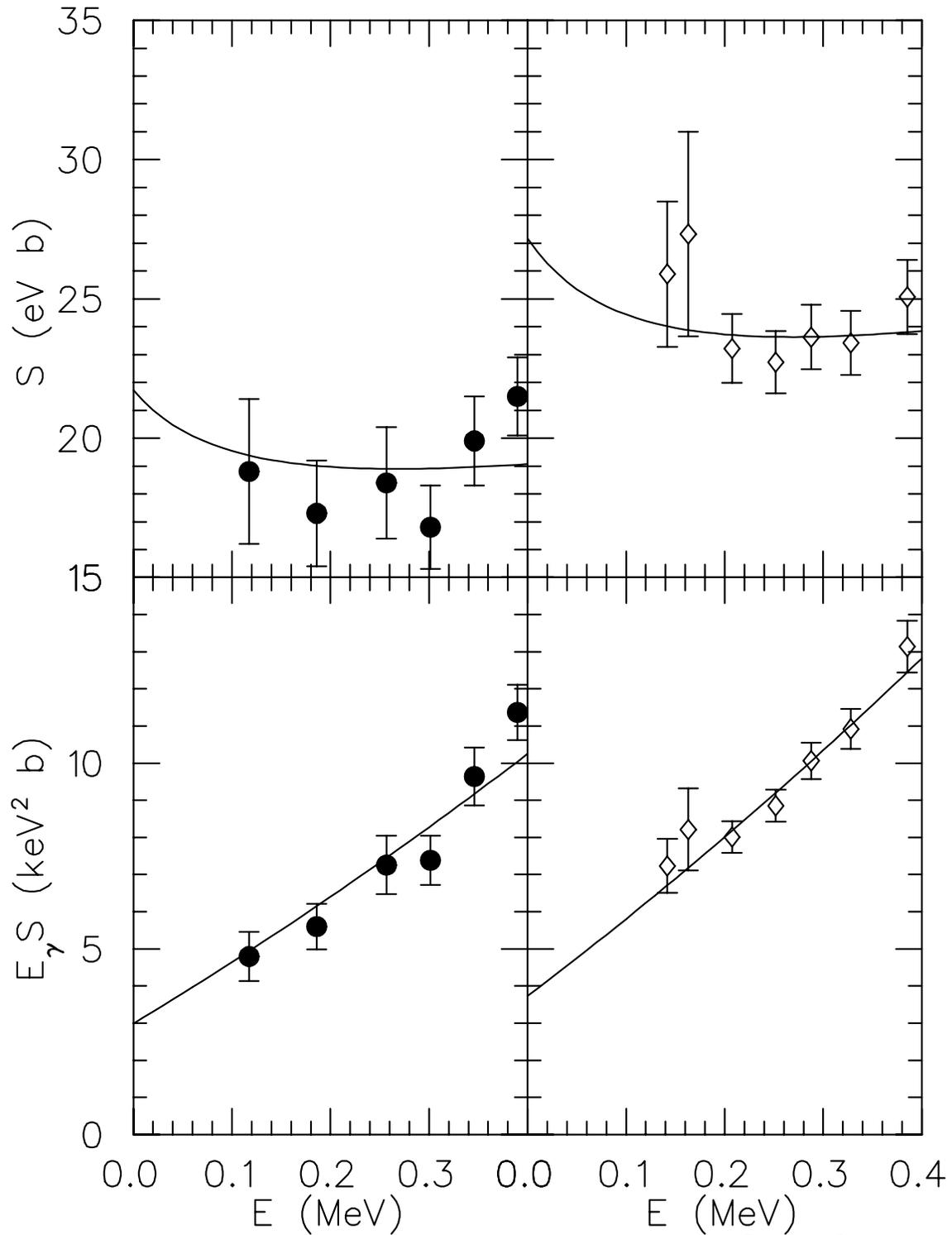,width=15cm}
\caption[]{The low-energy part of the astrophysical $S$ factor and $E_\gamma S$
for $^7$Be($p,\gamma$)$^8$B. The data of Filippone \cite{Filippone}
(circles) and Kavanagh \cite{kavanagh} (diamonds) are compared to the
results of the calculations as described in the text (solid line).
}
\label{s17low}
\end{figure}

In Fig.~\ref{s17low}, the data of Filippone \cite{Filippone} (circles)
and Kavanagh \cite{kavanagh} (diamonds) for the $S$ factor (top
panels) and the product $E_\gamma S$ (bottom panels) for the
$^7$Be($p,\gamma$)$^8$B reaction are presented for energies well below
the $E=633$ keV M1 resonance.  We take the data as normalized by
Johnson {\em et al.}\cite{Calvin} to $\sigma_{dp}= 157$ mb.  The
curves are similar in form to Eq.~(\ref{form}), but with a quadratic
term added,
\begin{equation}
S = n \frac{1 + c_1 E + c_2 E^2}{E_\gamma}
  = n \frac{1 + c_1 E + c_2 E^2}{E + E_{\rm B}}.
\label{form2}
\end{equation}
The values of $n$ and $c_i$ for this reaction are also listed in
Table~\ref{param}. Different values of the normalization $n$ are
required to reproduce the Filippone and Kavanagh data, but the $c_i$
are determined from the threshold energy dependence of a direct-capture
calculation following Ref.~\cite{williams}. A cut-off radius of
$r_0=2.3$~fm was chosen to be consistent with the phase shift and energy
dependence found by Barker\cite{barker}.  The upturn at threshold is
clearly observed in the results of the calculation. The data are
insufficient to determine this behavior or, equivalently, $c_i$. It
will be very difficult to experimentally confirm this upturn since it
is only pronounced below 100 keV. Fortunately it is theoretically well
understood and both $n$ and $c_1$ depend primarily on the asymptotic
normalization, spectroscopic factor and properties of the Coulomb
force. Note that the curves presented in Fig.~\ref{s17low} should not
be mistaken for a serious attempt at determining the $S_{17}$ factor
at zero energy; rather, they are illustrative of the energy
dependence.

The straight line approximation for $E_{\gamma}S$ is valid for the
$^{16}$O($p,\gamma$)$^{17}$F$^{\ast}$ reaction up to $\approx
3$~MeV. However, the quadratic approximation for the $^7$Be($p,\gamma$)$^8$B
reaction is not valid for energies above 0.4~MeV.  Initially, the
breakdown is caused by the resonance at 0.633~MeV. Above the resonance
higher order terms in $E$, arising predominantly from $d$-wave
direct capture, become significant.

In conclusion we see that the threshold peak in the $S$ factor is
associated with weakly bound states and arises from a pole at
$E_\gamma=0$. Those bound states in $^8$B and $^{17}$F are halo in
nature and so the associated radial integrals are, by necessity, long range.

The authors would like to thank F.C.~Barker, L.~Buchmann, H.~Fearing,
T.~Hemmert, K.~Langanke and E.~Vogt for useful discussions and
R.~Morlock for making tables of his data available.  Financial
assistance from the Natural Sciences and Engineering Research Council
of Canada is gratefully acknowledged.

\begin{table}
\caption[]{The numerical constants $c_i$ and $E_{\rm B}$ used to determine
the energy dependences, with the normalizations $n$ used when displaying
the curves with the discussed data sets.}
\label{param}
\begin{tabular}{ccccc}
Reaction & 	$c_1$ (MeV$^{-1}$) &	$c_2$ (MeV$^{-2}$) & $E_{\rm
B}$ (MeV) & $n$ (keV$^2$b)\\  
\hline
$^{16}$O($p,\gamma$)$^{17}$F$^{*}$&
	 	1.18 & 			0		 & 0.1052 & 
 $1.59\times10^3$\\ 
$^7$Be$(p,\gamma)^8$B &
		5.36 & 			1.80 		 & 0.1375 & 3.74$^a$\\
	            & & 				 &        & 2.99$^b$ \\
\end{tabular}
\footnotesize{(a) Kavanagh {\em et al.} \cite{kavanagh}\\(b) Filippone
{\em et al.} \cite{Filippone} }
\end{table}

\end{document}